\documentclass[a4paper,11pt]{article}

\usepackage{jheppub}
\usepackage{graphicx,amsmath,amssymb,subfigure}

\newcommand{\eg}{\textit{e.g.}}

\begin{document}

\title{In-Medium Charmonium Production in Proton-Nucleus Collisions}

\author[a]{Xiaojian Du}
\author[a]{Ralf Rapp}
\affiliation[a]{Cyclotron Institute and Department of Physics and Astronomy,
Texas A\&M University, College Station, TX
  77843-3366, USA}

\emailAdd{xjdu@physics.tamu.edu}
\emailAdd{rapp@comp.tamu.edu}

\date{\today}

\renewcommand{\thefootnote}{\arabic{footnote}}
\setcounter{footnote}{0}

\begin{abstract}
{We study charmonium production in proton-nucleus ($p$-A) collisions focusing on final-state
effects caused by the formation of an expanding medium. Toward this end we utilize a rate
equation approach within a fireball model as previously employed for a wide range of
heavy-ion collisions, adapted to the small systems in $p$-A collisions. The initial geometry
of the fireball is taken from a Monte-Carlo event generator where
initial anisotropies are caused by fluctuations. We calculate the centrality and
transverse-momentum dependent nuclear modification factor ($R_{p{\rm A}}$) as
well as elliptic flow ($v_2$) for both $J/\psi$ and $\psi(2S)$ and compare them to experimental
data from RHIC and the LHC. While the $R_{p{\rm A}}$s show an overall fair agreement
with most of the data, the large $v_2$ values observed in $p$-Pb collisions at the LHC cannot
be accounted for in our approach. While the former finding generally supports the formation
of a near thermalized QCD medium in small systems, the discrepancy in the $v_2$ suggests
that its large observed values are unlikely to be due to the final-state collectivity of
the fireball alone.}
\end{abstract}


\maketitle

\section{Introduction}
\label{sec_introduction}
Quarkonia, bound states of a heavy quark ($Q$=$b,c$) and its antiquark, are pristine
objects to study the properties of the fundamental color force. The discovery of charmonia
and bottomonia in the 1970s led to the development of the Cornell potential~\cite{Eichten:1974af}
which remains valid to date. To study its modifications in QCD media, large-scale experimental
and theoretical efforts are ongoing to measure and interpret the production systematics of
quarkonia in ultra-relativistic heavy-ion collisions
(URHICs)~\cite{Rapp:2008tf,BraunMunzinger:2009ih,Kluberg:2009wc,Mocsy:2013syh}. An
interplay of suppression processes in a quark-gluon plasma (QGP)
and subsequent regeneration in the later stages of the fireball evolution is, in principle,
required to arrive at a comprehensive description of quarkonium observables in URHICs. In
particular, the significance of regeneration reactions has been validated by an excitation
function for the $J/\psi$ which transits from a regime of strong suppression at SPS and
RHIC energies to significantly less suppression at the LHC, with the extra yield mostly
concentrated at low transverse momenta ($p_T$) (cf. Ref.~\cite{Rapp:2017chc} for a recent
overview). However, when accounting for the so-called cold-nuclear matter (CNM) effects,
which, in particular, cause a large suppression as observed in proton-nucleus ($p$-A)
collisions at the SPS~\cite{Arnaldi:2010ky}, the excitation function becomes rather flat.
Already at that time, this demonstrated the importance of $p$-A collisions as a reference for
the effects in AA systems.

The role of small colliding  system ($p$-A/$d$-A collisions) has recently received renewed
interest at RHIC and the LHC, including the quarkonium
sector~\cite{Adare:2013ezl,Adare:2016psx,Sirunyan:2017mzd,Aaboud:2017cif,Sirunyan:2018pse,Aaboud:2018quy,Adam:2015jsa,Leoncino:2016xwa,Adam:2016ohd,alice:ALICE-PUBLIC-2017-007,Adam:2015iga,Abelev:2014zpa,Acharya:2018kxc,Aaij:2017cqq,Acharya:2017tfn,CMS:2018xac}.
A moderate $J/\psi$ suppression (enhancement) has been found in both $d$-Au collisions at RHIC and forward (backward) rapidity $p$-Pb
collisions at the LHC, largely consistent with CNM effects (most notably a nuclear anti-/shadowing
of the initial parton distribution
functions)~\cite{Liu:2013via,Arleo:2014oha,Ferreiro:2014bia,Chen:2016dke,Ducloue:2016pqr,Albacete:2017qng}.
However, indications for a much stronger suppression of the $\psi(2S)$ have been observed
in $d$-Au collisions at RHIC and more precisely established at the LHC, especially at backward
rapidity (the nucleus-going direction) where the light-hadron multiplicity is the highest.
These observations have been explained by final-state absorption on
comovers~\cite{Ferreiro:2014bia}, or, closely relate, by dissociation reactions in the
QGP and hadronic phase of a fireball formed in these reactions~\cite{Du:2015wha,Chen:2016dke}
(both comover and thermal-suppression approaches feature comparable dissociation cross
sections as well as fireball energy densities and timescales).

In the present paper we expand on our earlier results for $d$-Au collisions~\cite{Du:2015wha},
by extending the kinetic rate-equation framework to $p$-Pb collisions at the
LHC, including the calculation of $p_T$ spectra and rapidity dependencies. We construct an
anisotropically expanding fireball based on initial asymmetries taken from Glauber model
estimates of initial-shape fluctuations~\cite{Nagle:2018}, which also allows us to compute
charmonium elliptic flow. We recall that the rate equation approach necessarily includes
regeneration contributions, which occur even in the presence of a single $c\bar c$ pair
(sometimes referred to as ``diagonal" regeneration or canonical limit). Their significance in $p$-Pb collisions has been suggested, \eg, in Ref.~\cite{Liu:2013via}.

Our paper is organized as follows.
In Sec.~\ref{sec_rate}, we summarize the main components of the kinetic rate-equation/transport
model developed for AA collisions and describe its extension to $p$-A collisions, in particular
the anisotropic fireball evolution. In Sec.~\ref{sec_raa}, we discuss our theoretical
results for the nuclear modification factor as a function of centrality and $p_T$, by first
revisiting the $d$-Au system at RHIC, followed by 5.02\,TeV and 8.16\,TeV $p$-Pb collisions at
the LHC, and compare to available experimental data from PHENIX, ALICE and LHCb.
In Sec.~\ref{sec_v2}, we discuss the $v_2$ results from our model in
comparison to 8.16\,TeV ALICE and CMS data.
In Sec.~\ref{sec_conclusion}, we summarize and conclude.

Our definition of backward (forward) rapidity in a $p$/$d$-A collisions follows the
experimental convention of referring to the nucleus-going (proton-going) direction.

\section{Transport Approach to Proton-Nucleus Collisions}
\label{sec_rate}
The kinetic-rate equation approach developed in Refs.~\cite{Grandchamp:2003uw,Zhao:2010nk}
is based on a space-momentum integrated Boltzmann equation,
\begin{equation}
\frac{\mathrm{d}N_\Psi(\tau)}{\mathrm{d}\tau}=-\Gamma_\Psi(T(\tau))\left[N_\Psi(\tau)-
N_\Psi^{\rm eq}(T(\tau))\right]
\label{rate_eq}
\end{equation}
where $N_\Psi(\tau)$ denotes the time-dependent number of charmonium states
$\Psi=J/\psi,\psi(2S),\chi_c$(1P). The transport parameters are the inelastic reaction
rate $\Gamma_\Psi$ and the equilibrium limit (controlling the regeneration contribution),
\begin{equation}
N_\Psi^{\rm eq}(T) = V_{\rm FB} \gamma_c^2 d_\Psi \int \frac{d^3p}{(2\pi)^3} f_{\Psi}^{\rm eq}(E_p;T)
\end{equation}
at a given temperature, $T$, of the ambient medium; $V_{\rm FB}=V_{\rm FB}(\tau)$ denotes the time-dependent
fireball volume which we parameterize in Eq.(\ref{VFB}) below, and $f_{\Psi}^{\rm eq}$ the
thermal Bose distribution of the $\Psi$-state with degeneracy $d_\Psi$.
The charm-quark fugacity, $\gamma_c$, is determined
by charm-quark conservation in the fireball,
\begin{equation}
N_{c\bar{c}}=\frac{1}{2}\gamma_c n_{\rm op} V_{\rm FB}\frac{I_1(\gamma_c n_{\rm op}
V_{\rm co})}{I_0(\gamma n_{\rm op} V_{\rm co})}+\gamma_c^2 n_{\rm hid} V_{\rm FB} \ ,
\end{equation}
where $n_{\rm op(hid)}=n_{\rm op(hid)}(T)$ denotes the  density of all open (hidden) charm states in
the system, and $N_{c\bar{c}}$ is the total number of charm-quark pairs in the fireball
which is calculated from the $c\bar{c}$ production cross section in $pp$ ($\sigma_{c\bar{c}}=\frac{\mathrm{d}\sigma_{c\bar{c}}}{\mathrm{d}y}\Delta y$ with $\Delta y\simeq1.8$ for one fireball) at given collision
energy $N_{c\bar{c}}=\frac{\sigma_{c\bar{c}}}{\sigma_{\rm inel}}N_{\rm coll}$, where $N_{c\bar{c}}$ is the number of binary collisions at a given centrality of a $p$-A (or $d$-A) collision, and $\sigma_{\rm inel}$ is the total inelastic cross section. The $c\bar{c}$ pair correlation volume, $V_{\rm co}=\frac{4}{3}(R_{c\bar{c}}+v_{c\bar{c}}\tau)^3$
(with initial $c\bar{c}$ pair radius $R_{c\bar{c}}$=1.2\,fm and $c\bar{c}$ relative velocity $v_{c\bar{c}}$=0.6), represents the sub-volume that a pair can
occupy after its production (sub-volumes of multiple pairs are merged once they overlap).
Regeneration is only active for temperatures below the respective dissociation temperatures
of each state $\Psi$ (which are $T_c$, 1.3\,$T_c$ and 2~$T_c$ for the $\psi(2S)$, $\chi_c$
and $J/\psi$, respectively~\cite{Zhao:2010nk}).
To account for incomplete charm-quark thermalization~\cite{Grandchamp:2002wp}, the
equilibrium limit has been corrected by a thermal relaxation time factor,
\begin{equation}
R(\tau)=1-{\rm exp}\left(-\int^{\tau}_{\tau_0}\frac{\mathrm{d}\tau'}{\tau_c}\right) \ ,
\end{equation}
where $\tau_c$ is the thermal relaxation time of charm quarks, assumed to be
constant~\cite{Riek:2010fk}) at $\tau_c$=4\,fm/c.
Langevin simulations of charm quarks in $p$-Pb collisions~\cite{He:2014cla} have
found that a significantly larger relaxation time, by a factor of 3-5, is necessary to
be compatible with the observed $R_{p {\rm A}}$ of $D$-mesons; we therefore increase our
previously employed relaxation time from 4 to 15~fm.
The reaction rate $\Gamma_\Psi$ accounts for quasifree
dissociation in the QGP and an extended set of dissociation reactions in the hadronic
phase as constructed in Ref.~\cite{Du:2015wha} based on $SU$(4) meson
exchange~\cite{Lin:1999ad}, supplemented by phase-space considerations for higher resonance states.
Similar to Ref.~\cite{Du:2015wha}, we increase the QGP dissociation rate of the loosely
bound (or even unbound) $\psi(2S)$ by a factor of 3 to account for non-perturbative effects
on heavy-quark interactions in the QGP at moderate temperatures~\cite{Liu:2017qah}.
Uncertainties of the hadronic $\psi(2S)$ dissociation rate are accounted for by increasing
the baseline rate by up to a factor of 2. However, the variation of the hadronic
rate has very little impact on the final $R_{p {\rm A}}$, due to a near compensation of
dissociation and regeneration contributions.

To schematically account for the effects of quantum evolution in the early stages of the
charmonium evolution, we utilize formation times, $\tau_\Psi^{\rm form}$, for the different
states ($J/\psi$, $\psi(2S)$, $\chi_c(1P)$) that are assumed to have a range of 1-2\,fm
to reflect uncertainties associated with their binding energies. Their effect is rather
small in semi-/central AA collisions, but becomes augmented in small systems due to shorter fireball lifetimes. The formation times not only influence (suppress) the magnitude of the
early charmonium dissociation but also modify its $p_T$ dependence due to Lorentz time
dilation implemented via
\begin{equation}
\tilde{\Gamma}_{\Psi}(\vec{p}_T,T(\tau))=\Gamma_{\Psi}(\vec{p}_T,T(\tau))\frac{\tau}{\tau_{\Psi}^{\rm form}}\frac{m_{\Psi}}{\sqrt{p_T^2+m_{\Psi}^2}} \
\end{equation}
for $\tau<\tau_{\Psi}^{\rm form}\frac{\sqrt{p_T^2+m_{\Psi}^2}}{m_{\Psi}}$.

The time evolution of the fireball volume and temperature is constructed through an
isotropic expansion with conserved total entropy,
\begin{equation}
S_{\rm tot}=s(T)V_{\rm FB}(\tau) \ ,
\end{equation}
matched to the experimentally measured charged-hadron multiplicity in a given rapidity
region of a nuclear collision system. The entropy density, $s(T)$, is calculated for a
quasiparticle QGP and hadron resonance gas with a mixed phase at $T_c$=180\,MeV (in
recent work~\cite{Du:2017qkv} we have found that the use of a more modern lQCD-based EoS
affects the temperature dependence of the fireball cooling (and thus the bottomonium
kinetics) rather little; to keep consistency with our published charmonium results,
we defer the EoS update to a future work).

\begin{figure}[!t]
\centering
\includegraphics[width=8cm]{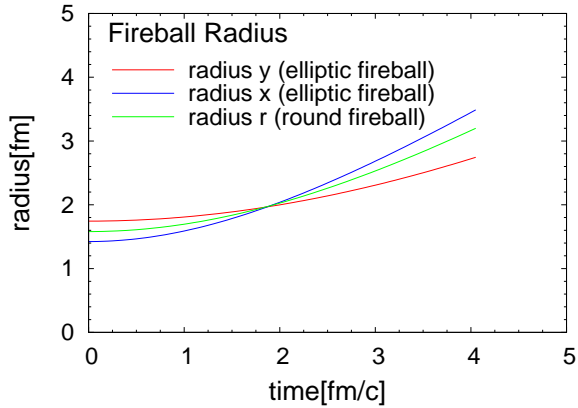}
\caption{Transverse radii for the expanding fireball in round (green line) and elliptic (red
and blue lines) geometry for central $p$-Pb(5.02\,TeV) collisions with  $a_T$=0.24\,/fm for the
round fireball, and $a_{x,y}$=0.34,0.13\,/fm for the elliptic fireball.}
\label{fig_fb}
\end{figure}

Here, we extend our previously used cylindrical volume expansion to allow for a
elliptic deformation in the transverse plane,
\begin{equation}
V_{\rm FB}=\left(z_0+v_z \tau\right)\pi R_x(\tau) R_y(\tau) \ ,
\label{VFB}
\end{equation}
where $z_0$ is the initial longitudinal size related to the formation time via
$\tau_0= z_0/ \Delta y$, which we assume to be 0.8\,fm, somewhat larger than in AA
collisions to account for the reduced overlap density in the smaller $p$-A systems.
The transverse radii in $x$- and $y$-direction are parameterized as
\begin{eqnarray}
R_x(\tau)&=& R_0-d+\frac{\sqrt{1+(a_x\tau)^2}-1}{a_x}\\
R_y(\tau)&=& R_0+d+\frac{\sqrt{1+(a_y\tau)^2}-1}{a_y} \ ,
\end{eqnarray}
with $R_0=\frac{R_x^0+R_y^0}{2}$ and  $d=\frac{R_y^0-R_x^0}{2}$ .
The initial radii $R_x^0$ and $R_y^0$ are estimated from the eccentricity of the initial
distribution of a Monte-Carlo Glauber event generator~\cite{Nagle:2018},
\begin{equation}
e=\frac{(R_y^0)^2-(R_x^0)^2}{(R_y^0)^2+(R_x^0)^2}=0.2 \ ,
\label{eccentricity}
\end{equation}
with an initial transverse area $A_\perp^{p{\rm Pb}}$=$\pi R_ x^0 R_y^0$=7.8\,fm$^2$~\cite{Nagle:2018}.
The surface velocities of the fireball are
computed with a relativistic acceleration ansatz
\begin{eqnarray}
v_x(\tau)&=& \frac{a_x\tau}{\sqrt{1+(a_x\tau)^2}}\\
v_y(\tau)&=& \frac{a_y\tau}{\sqrt{1+(a_y\tau)^2}}\ .
\end{eqnarray}
The parameters $a_x$=0.34/fm and $a_y$=0.13/fm are fixed in order to describe the
light-hadron (pion, kaon, proton) $p_T$ spectra and $v_2$ including their mass splitting
via the anisotropic blastwave formula, Eq.(\ref{blastwave}), at thermal freezeout
($\Delta a_T$=$\frac{a_x-a_y}{2}$ controls the magnitude of the $v_2$ and
$a_T$=$\frac{a_x+a_y}{2}$ its mass splitting).
The average transverse acceleration of $a_T$$\sim$0.24/fm, relative to our default
value of 0.1/fm in AA collisions, reflects the larger pressure gradients in $p$-A
collisions~\cite{Kalaydzhyan:2014lja}.
We have checked that our total $R_{p {\rm A}}$ results are
rather insensitive to this value over a range of accelerations, $a_T$=0.1-0.4/fm.
Larger accelerations slightly reduce both the suppression and regeneration
(compensating each other) due to shorter fireball lifetimes (most of the hot-matter
effects happen at relatively early times where longitudinal expansion dominates).
Blastwave fits of light-hadron spectra~\cite{Vazquez:2017jce} extract average transverse
velocities of up to $\sim$0.5 which would indeed require an transverse acceleration closer
to $\sim$0.4/fm in our fireball framework. However, the blastwave fits in $p$-Pb collisions
might be more sensitive, relative to AA collisions, to primordial hard components leaking
into the fit range; thus a smaller transverse acceleration might be preferred. We therefore
work with the values specified above unless otherwise stated.
We also note that with the current fireball parameterization, the radii $R_x$ and $R_y$ cross
at $\tau$$\sim$2\,fm, implying a transition to an in-plane deformation, cf.~Fig.~\ref{fig_fb}.
We have checked using different ans\"atze that this is a robust feature dictated by a rapid
build-up of the $v_2$ while approximately recovering light-hadron $v_2$ data in $p$-A collisions.
After the crossing, the in-plane acceleration should become smaller than out-of-plane.
While this feature is not explicitly guaranteed by our parameterization, the chosen
parameter values lead to a sign flip of the anisotropic component of the acceleration,
$\Delta a(\tau)=\frac{{\mathrm d}\Delta v(\tau)}{{\mathrm d}\tau}=\frac{1}{2}\left(\frac{{\mathrm d}v_x(\tau)}{{\mathrm d}\tau}-\frac{{\mathrm d}v_y(\tau)}{{\mathrm d}\tau}\right)$,
close to the crossing point of the radii. In any case, the net change in $v_2$ after this
point is small and has very little bearing on our results.

\begin{figure}[!t]
\begin{minipage}[b]{0.48\linewidth}
\centering
\includegraphics[width=1.12\textwidth]{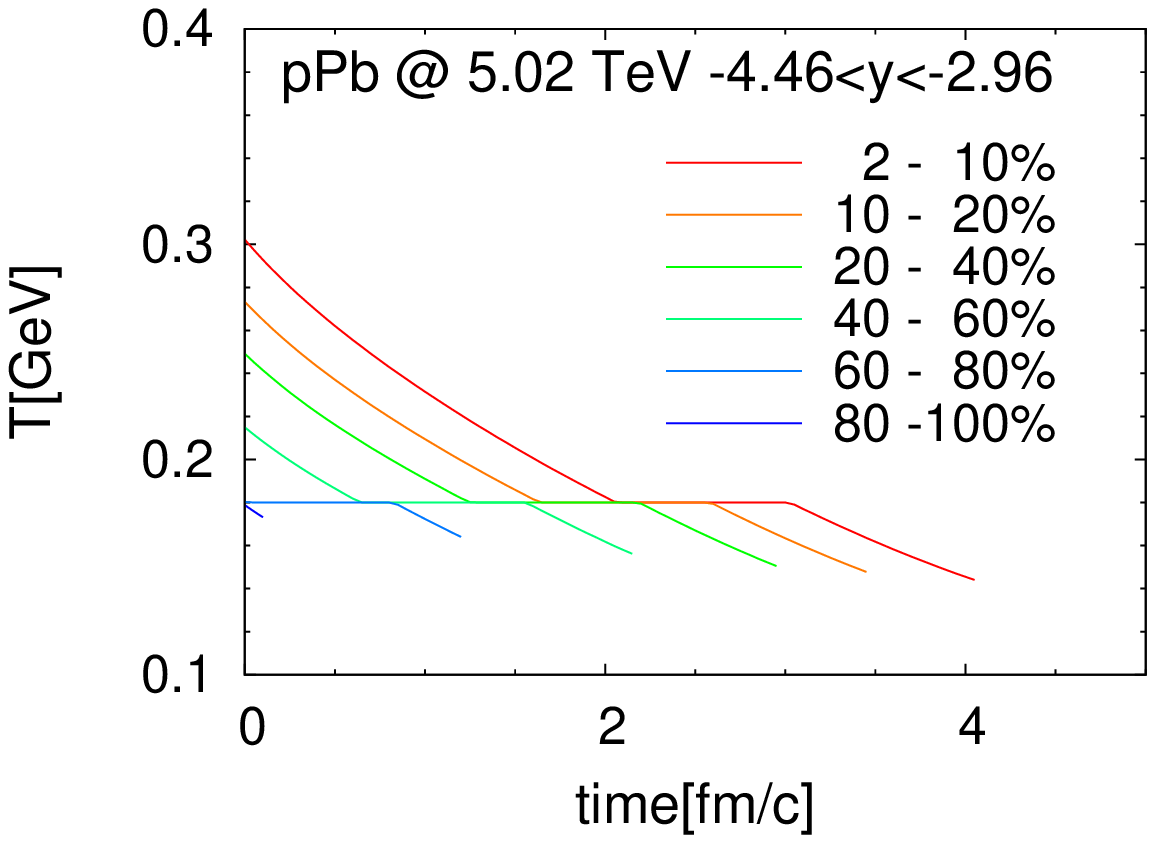}
\end{minipage}
\hspace{\fill}
\begin{minipage}[b]{0.48\linewidth}
\centering
\includegraphics[width=1.12\textwidth]{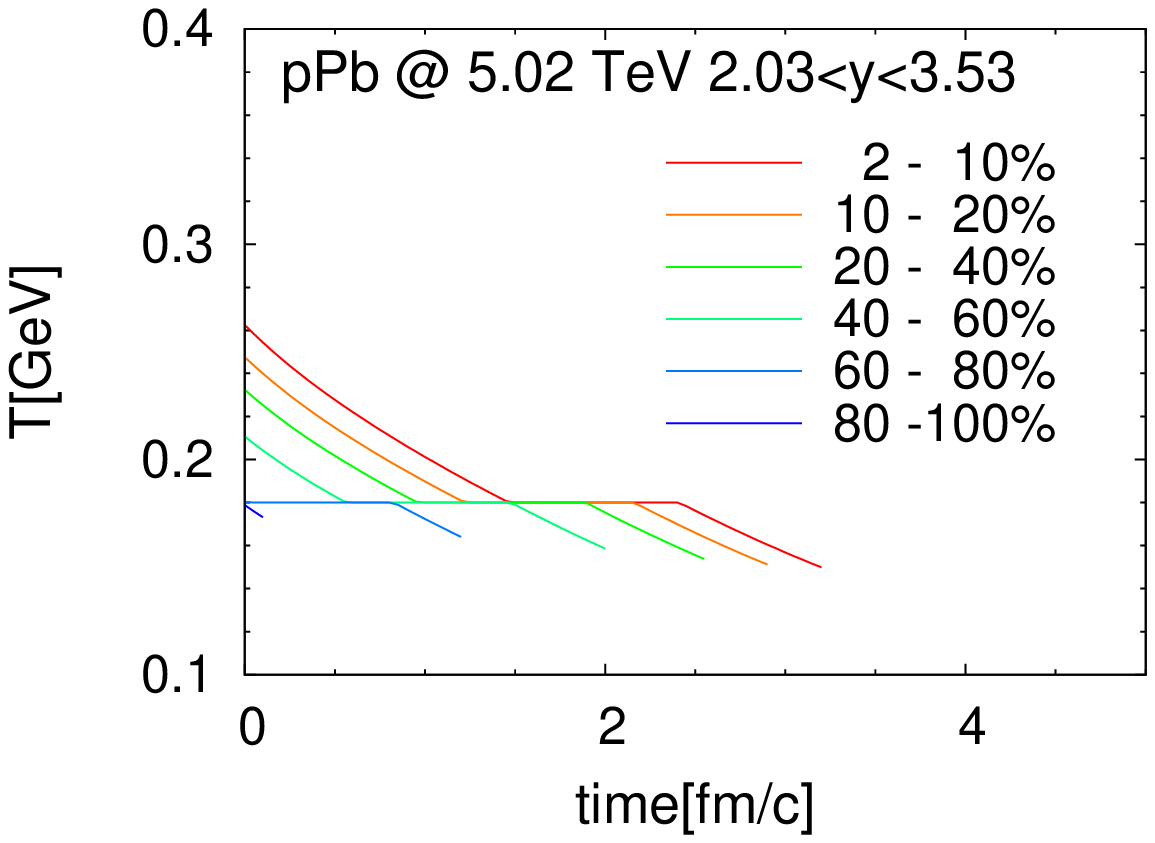}
\end{minipage}

\vspace*{0.5cm}
\begin{minipage}[b]{0.48\linewidth}
\centering
\includegraphics[width=1.12\textwidth]{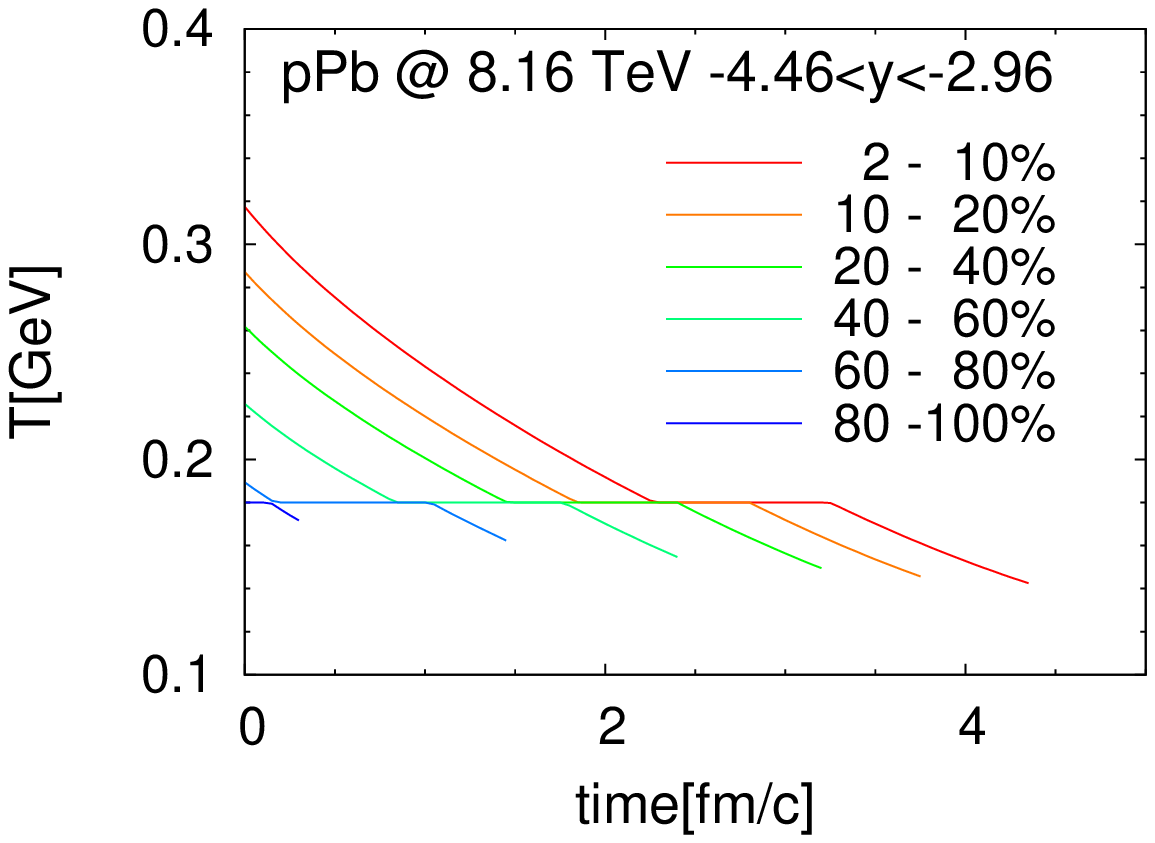}
\end{minipage}
\hspace{\fill}
\begin{minipage}[b]{0.48\linewidth}
\centering
\includegraphics[width=1.12\textwidth]{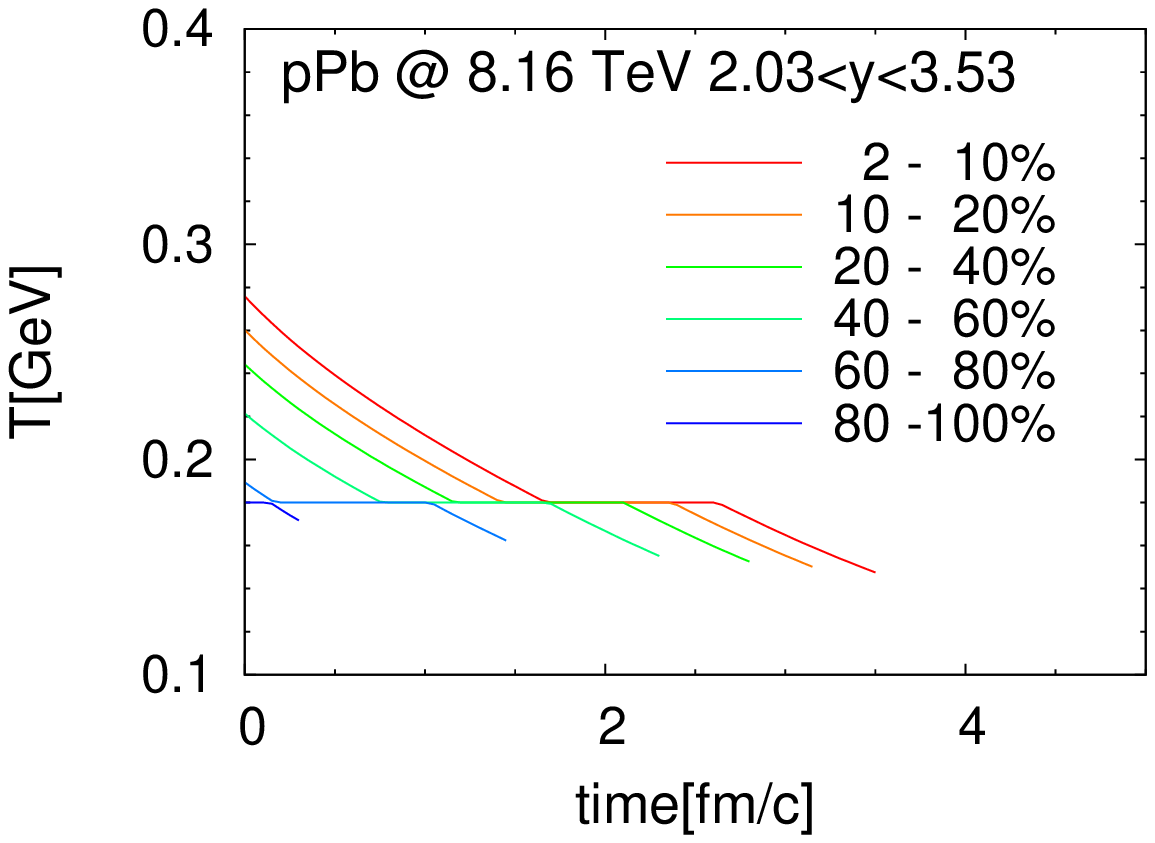}
\end{minipage}
\caption{Temperature evolution from the elliptic fireball model  for
different centralities in $p$-Pb collisions at 5.02\,TeV (upper panels) and 8.16\,TeV
(lower panels) at backward (left column) and forward (right column) rapidities.}
\label{fig_tevo}
\end{figure}

Much like for different centralities in AA collisions, one can expect significant
variations in the kinetic-freezeout temperature as a function of multiplicity in small
systems: for a smaller total entropy the criterion that the mean-free-path is comparable
to the fireball size (or inverse expansion rate) is reached at a larger particle density
(or temperature). Guided by Ref.~\cite{Vazquez:2017jce} we implement this effect by a
centrality-dependent freezeout temperature as
\begin{equation}
T_{\rm fo}(N_{\rm ch})=145~{\rm MeV}\left(\frac{S_{\rm tot}(N_{\rm ch})}{552}\right)^{-\frac{1}{12}}
\ .
\end{equation}
The temperature evolutions following from this construction are summarized in
Fig.~\ref{fig_tevo} for 5.02\,TeV and 8.16\,TeV $p$-Pb collisions for different ``centralities"
(or rather, $N_{\rm ch}$) at forward and backward rapidities.

To compute $p_T$ spectra within our approach, we decompose the rate equation into a
solution for the primordial suppressed component, $N_{\rm prim}$, and the regeneration
component, $N_{\rm reg}$. For the former, we solve a Boltzmann equation,
\begin{equation}
\frac{\partial f_{\rm prim}(\vec{x}_T,\vec{p}_T,\tau)}{\partial\tau}
+\vec{v}_\Psi\cdot\frac{\partial f_{\rm prim}(\vec{x}_T,\vec{p}_T,\tau)}{\partial\vec{x}_T}
=-\Gamma_\Psi(\vec{p}_T,T(\tau))f_{\rm prim}(\vec{x}_T,\vec{p}_T,\tau)
\label{bolt_eq}
\end{equation}
where $\vec{v}_\Psi$ denotes the charmonium velocity in the lab frame.
The 2-dimensional vectors $\vec{p}_T(p_T,\theta_p)$  and $\vec{x}_T(r,\theta_r)$ encode
anisotropies in the transverse-momentum and coordinate plane, respectively,
originating from different path lengths when traversing the elliptic fireball.
The dissociation rate $\Gamma_\Psi(\vec{p}_T,T(\tau))$ is evaluated in the medium
rest frame where temperature is defined. In principle, there is a difference between
the proper time $\tau$ in the rate equation (Eq.(\ref{rate_eq})) and the longitudinal
proper time $\tau$=$\sqrt{t^2-z^2}$ in the Boltzmann equation (Eq.(\ref{bolt_eq})).
However, since we place a single fireball at the appropriate rapidity for a given
experiment, we do not correct its thermal rapidity width for time dilation.
We also neglect the Lorentz contraction effects in the transverse volume expansion when
solving the Boltzmann equation. The maximal effect at the surface with $v_T\sim$0.5 is
less than 15\%, which is well within the uncertainty range of our acceleration parameter
of 0.1-0.4\,/fm.

For the regeneration yield in the elliptic fireball, we use the yield
obtained from the rate equation and
approximate its $p_T$-spectrum by an anisotropic blastwave
description~\cite{Huovinen:2001cy,Retiere:2003kf},
\begin{equation}
\frac{\mathrm{d}N_{\rm reg}}{p_T\mathrm{d}p_T\mathrm{d}\theta_p}\propto m_T
\int\limits^{2\pi}_0\int\limits^{R_{\rm max}(\theta_r)}_0 \mathrm{d}\theta_r r\mathrm{d}r
K_1\left(\beta\right) {\rm e}^{\alpha{\rm cos}(\theta_p-\theta_b)}
\label{blastwave}
\end{equation}
with $\alpha$=$\frac{p_T}{T} \sinh\rho(r,\theta_r,\tau)$, $\beta$=$\frac{m_T}{T} \cosh\rho(r,\theta_r,\tau)$
and transverse mass $m_T$=$\sqrt{p_T^2+m_\Psi^2}$. The transverse-flow rapidity,
$\rho(r,\theta_r,\tau)$=${\tanh^{-1}}(v_\perp(r,\theta_r,\tau))$, is evaluated in terms of
the fireball expansion velocity profile,
$v_\perp(r,\theta_r,\tau)=(r/R_{\rm max}(\theta_r,\tau))v_s(\theta_r,\tau)$,
with surface velocity
\begin{equation}
v_s(\theta_r,\tau)=\frac{v_x(\tau)+v_y(\tau)}{2}
+\frac{v_x(\tau)-v_y(\tau)}{2}{\rm cos}(2\theta_b(\theta_r,\tau)) \ .
\end{equation}
Since the surface velocity on the semi-minor axis ($x$-direction, with $\theta_b$=$\theta_r$=0) is larger than on the semi-major axis ($y$-direction, $\theta_b$=$\theta_r$=$\pi$/2),
it generates a positive $v_2$.
The surface radius
\begin{equation}
R_{\rm max}(\theta_r,\tau)=\frac{1}
{\sqrt{\left(\frac{{\rm sin}(\theta_r)}{R_y(\tau)}\right)^2
+\left(\frac{{\rm cos}(\theta_r)}{R_x(\tau)}\right)^2}}
\label{r_bound}
\end{equation}
depends on the coordinate angle $\theta_r$ and represents the boundary of
the fireball, while
\begin{equation}
\theta_b(\theta_r,\tau)={\rm arctan}\left(\left(\frac{R_x(\tau)}{R_y(\tau)}\right)^2 {\rm tan}(\theta_r)\right)
\end{equation}
characterizes the direction of the medium flow perpendicular to the fireball boundary.

To compute the denominator of the $p_T$-dependent nuclear modification factor,
and as an initial condition to the Boltzmann equation, we need the initial
charmonium phase space distributions.
We assume a factorization into transverse-momentum and coordinate space. For the $p_T$
distribution we employ an ALICE parametrization~\cite{Bossu:2011qe,Book:2015} of the
spectra in $pp$ collisions of the form
\begin{equation}
\frac{\mathrm{d}N_{pp}}{p_T\mathrm{d}p_T}(p_T)=f_{pp}(p_T) \propto \left(1+\left(\frac{p_T}{B}\right)^2\right)^{-A} \ ,
\end{equation}
with $A$=3.73(3.70), $B$=3.81(5.10) for $J/\psi$ ($\psi(2S)$).
The initial coordinate distribution is assumed to be a Gaussian,
\begin{equation}
f(\vec{x}(r,\theta_r))=f_0 {\rm exp}\left(-\frac{(r{\rm sin}(\theta_r))^2}{(R_y^0)^2}
-\frac{(r{\rm cos}(\theta_r))^2}{(R_x^0)^2}\right) \ ,
\end{equation}
so that on the initial fireball boundary $f={f_0}/{\rm e}$, and the enclosed elliptic
area is equal to the initial transverse area ($\pi R_x^0 R_y^0$) within which all initial
$c\bar c$ pairs are assumed to be produced (controlled by the normalization $f_0$).

The CNM effects are implemented in two steps. We first estimate the magnitude of the reduction
(or enhancement) of the $c\bar{c}$ and $\Psi$ yields from (anti-) shadowing using the
EPS09-LO and EPS-NLO framework~\cite{Eskola:2009uj,Vogt:2010aa,Vogt:2015uba} at given rapidity and collision energy
(defining an error band encoded in our final results). We then approximate the $p_T$ dependence of the
CNM effects by a Gaussian broadening to represent both the original Cronin effect as well as the
$p_T$-dependence of shadowing,
\begin{equation}
\frac{\mathrm{d}N_{p{\rm A}}}{p_T\mathrm{d}p_T}(p_T,b)=
\int\mathrm{d}^2p_T^\prime\frac{\exp(-\frac{{p_T^\prime}^2}{a_{gN}L(b)})}{\pi a_{gN}L(b)}
\frac{\mathrm{d}N_{pp}}{p_T\mathrm{d}p_T}(|\vec{p}_T-\vec{p}_T^{\,\prime}|) \
\end{equation}
where $a_{gN}$ is the broadening per unit path length and $L(b)$ the mean path length of
the gluon in $p$/$d$-A collisions before fusing into a $\Psi$ state~\cite{Hufner:2001tg}.
It can be calculated as:
\begin{equation}
L(b)=\frac{\int \mathrm{d}^2r\mathrm{d}z_A\mathrm{d}z_B
\left(l^A(r_A,z_A)+l^B(r_B,z_B)\right)
K(b,r,z_A,z_B)}{\int \mathrm{d}^2r\mathrm{d}z_A\mathrm{d}z_B K(b,r,z_A,z_B)}
\end{equation}
for a nucleus $A$ and proton/deuteron $B$,
$r_A=|\vec{r}-\frac{\vec{b}}{2}|$ and $r_B=|\vec{r}+\frac{\vec{b}}{2}|$. In the above
expression, $l^A(r_A,z_A)=\frac{A}{\rho_0}\int_{-\infty}^{z_A}\mathrm{d}z\rho_A(r_A,z)$,
$l^B(r_B,z_B)=\frac{B}{\rho_0}\int_{z_B}^{\infty}\mathrm{d}z\rho_B(r_B,z)$ with
$\rho_{\rm A}(r_A,z_A)=\rho_{\rm A}(\sqrt{r_A^2+z_A^2})$ the Woods-Saxon distribution.
The kernel
\begin{eqnarray}
K(b,r,z_A,z_B)=\rho_A(r_A,z_A)\rho_B(r_A,z_A)
\nonumber
{\rm e}^{-\sigma_{\rm abs}\left[(A-1)
\int_{-\infty}^{z_A}\mathrm{d}z\rho_A(r_A,z)+(B-1)\int_{z_B}^{\infty}
\mathrm{d}z\rho_B(r_B,z)\right]}
\nonumber\\
\end{eqnarray}
represents the coordinate distribution of partons in the collision.
The path length in the proton and deuteron can be neglected, $l^B(r_B,z_B)\simeq 0$;
thus, only the size of nucleus $A$ contributes to the mean path length.  Treating the proton as a $\delta$-function,
$\rho_B(r_B,z_B)=\frac{2}{\pi}\delta(r_B^2)\delta(z_B)$, and with $B$=1,
$\vec{r}=-\frac{\vec{b}}{2}$ and $r_A=b$, the effective path length simplifies to
\begin{equation}
L(b)=\frac{\int\mathrm{d}z_A l^A(b,z_A)\rho_A(b,z_A)
{\rm e}^{-\sigma_{\rm abs}(A-1)\int_{-\infty}^{z_A}\mathrm{d}z\rho_A(b,z)}}
{\int\mathrm{d}z_A\rho_A(b,z_A) {\rm e}^{-\sigma_{\rm abs}(A-1)
\int_{-\infty}^{z_A}\mathrm{d}z\rho_A(b,z)}}
 \ .
\end{equation}
In the limit of zero absorption, it can be further simplified as
\begin{equation}
L(b)=\frac{A}{\rho_0}\frac{\int_{-\infty}^{\infty}\mathrm{d}z\int_{-\infty}^{z}\mathrm{d}z'\rho_A(b,z')\rho_A(b,z)}{\int_{-\infty}^{\infty}\mathrm{d}z\rho_A(b,z)}
\end{equation}
which is used for the evaluation of the $p_T$ broadening.
We associate the CNM effects for the $p_T$ dependence at the LHC with an effective broadening
parameter of $a_{gN}$=0.1-0.2\,GeV$^2$/fm reflecting the EPS09-LO vs.~NLO uncertainty at backward
rapidity, and $a_{gN}$=0.2-0.4\,GeV$^2$/fm to represent the steeper trend and uncertainty
from CGC calculations
at forward rapidity~\cite{Ducloue:2015gfa,Ducloue:2016pqr}. At mid-rapidity, we take an
intermediate range of $a_{gN}$=0.1-0.3\,GeV$^2$/fm.

The elliptic fireball allows the investigation of momentum anisotropies from final-state
interactions. After obtaining the anisotropic spectra,
$\mathrm{d}N_{\rm AA}/{\mathrm{d}^2p_T}(p_T,\theta_p)$, from the
primordial and regeneration components, the elliptic flow coefficient is readily calculated
as
\begin{equation}
v_2(p_T)=\frac{ \int\limits^{2\pi}_0
\frac{ \mathrm{d}N_{\rm AA} }{\mathrm{d}^2p_T}(p_T,\theta_p)
\cos(2\theta_p)\mathrm{d}\theta_p  }
{\int\limits^{2\pi}_0 \frac{\mathrm{d}N_{\rm AA}}{\mathrm{d}^2p_T}(p_T,\theta_p)
\mathrm{d}\theta_p} \ .
\end{equation}

\section{Nuclear Modification Factors for $J/\psi$ and $\psi(2S)$}
\label{sec_raa}
We are now in position to calculate the nuclear modification factors for charmonia in
$d$-Au(0.2\,TeV) collisions at RHIC (Sec.~\ref{ssec_rhic}) and in $p$-Pb(5.02,8.16\,TeV)
collisions at the LHC (Sec.~\ref{ssec_lhc}).
The cross section inputs will be specified in the respective sections.

\subsection{Deuteron-Gold Collisions at RHIC}
\label{ssec_rhic}
Compared to our previous studies of $d$-Au collisions at RHIC~\cite{Du:2015wha}, we here
implement the updates as described in the previous section to ensure consistency with
the new developments for $p$-Pb collisions. In particular,
the fireball is extended to elliptic geometry, and has a smaller initial transverse area
guided by the updates for $p$-Pb at the LHC described above; with a deuteron
size approximately twice the proton size, and an inelastic $NN$ cross section at RHIC of
2/3 of that at the LHC, we have
$A_\perp^{d{\rm Au}}= A_\perp^{p{\rm Pb}} \cdot 2 \cdot 2/3$=\,10.4\,fm$^2$.
As a result, the initial temperature in central $d$-Au now reaches $T_0\simeq$~245\,MeV.
While this increases the hot-matter suppression, it slightly enhances the escape effect
counter-acting the former. We also include regeneration contributions (neglected in
Ref.~\cite{Du:2015wha}) which contribute up to $\sim$0.05 at the $R_{d{\rm A}}$ level
and also counter-act the increased hot-matter suppression.

The input cross sections remain unchanged, with
$\frac{\mathrm{d}\sigma_{J/\psi}}{\mathrm{d}y}$=0.75\,$\mu$b~\cite{Adare:2011vq} for the $J/\psi$
and $\frac{\mathrm{d}\sigma_{c\bar{c}}}{\mathrm{d}y}$=123\,$\mu$b~\cite{Adare:2010de,Adamczyk:2012af} for
all $c\bar{c}$ pairs.
Cold-nuclear-matter effects are associated with EPS09 LO parton shadowing~\cite{Eskola:2009uj,Ferreiro:2014bia}
for both charmonium and open-charm production, whose centrality dependence we mimic by
using a nuclear absorption cross section of $\sigma_{\rm abs}$=0-2.4\,mb, as
before~\cite{Du:2015wha}.

\begin{figure}[!t]
\centering
\includegraphics[width=0.45\textwidth]{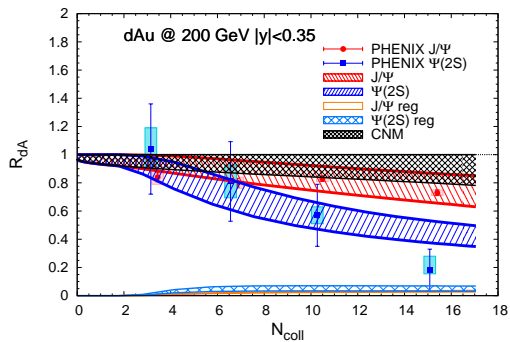}
\caption{Centrality dependent $R_{d{\rm A}}$ for $J/\psi$ (red bands) and $\psi(2S)$ (blue bands)
in 200\,GeV $d$-Au collision, compared with experimental
data~\cite{Adare:2013ezl}. The orange (light blue) band is for the $J/\psi$ ($\psi(2S)$)
regeneration component. The CNM effect only (black band) represents the uncertainty due
to shadowing (via an absorption cross section of 0-2.4\,mb) and is the major source of
uncertainty for the colored bands.}
\label{raa_dA}
\end{figure}
The model calculations are compared with PHENIX data~\cite{Adare:2013ezl} in Fig.~\ref{raa_dA}.
Fair agreement with experiment is found, very similar to our previous results~\cite{Du:2015wha}.

\subsection{Proton-Lead Collisions at the LHC}
\label{ssec_lhc}
In addition to the information specified in Sec.~\ref{sec_rate}, we here quote the input
cross sections as determined from $pp$ data at the LHC. For the $J/\psi$ we use
$\frac{\mathrm{d}\sigma}{\mathrm{d}y}$=3.0 and 3.6\,$\mu$b at backward (-4.46$<$$y$$<$2.96)
and forward (2.03$<$$y$$<$3.53) rapidity, respectively,
at 5.02\,TeV~\cite{Adam:2016rdg,Acharya:2017hjh}, and
$\frac{\mathrm{d}\sigma}{\mathrm{d}y}$=3.9(4.7)\,$\mu$b at backward (forward)
rapidity at 8.16\,TeV~\cite{Aaij:2017cqq}, and for $c\bar{c}$ pairs
$\frac{\mathrm{d}\sigma}{\mathrm{d}y}$=0.51(0.61)\,mb at backward (forward) rapidity
at 5.02\,TeV and $\frac{\mathrm{d}\sigma}{\mathrm{d}y}$=0.66(0.80)\,mb at 8.16\,TeV.
This amounts to a fixed $J/\psi$ over $c\bar{c}$ ratio of 0.58\,\% (as in our previous
work~\cite{Zhao:2011cv,Du:2015wha}). The charged-particle multiplicity determining the
total entropy of the fireball in the respective rapidity regions is extracted
from Refs.~\cite{Adam:2014qja,Spousta:2014xja} at 5.02\,TeV and guided by
Ref.~\cite{Sirunyan:2017vpr} for 8.16\,TeV.
In the following, we first discuss the centrality dependence (Sec.~\ref{sssec_centr})
and then the transverse-momentum dependence (Sec.~\ref{sssec_trans})
of $J/\psi$ and $\psi(2S)$ production in 5.02 and 8.16\,TeV $p$-Pb collisions.

\subsubsection{Centrality Dependence}
\label{sssec_centr}
\begin{figure}[!t]
\includegraphics[width=0.45\textwidth]{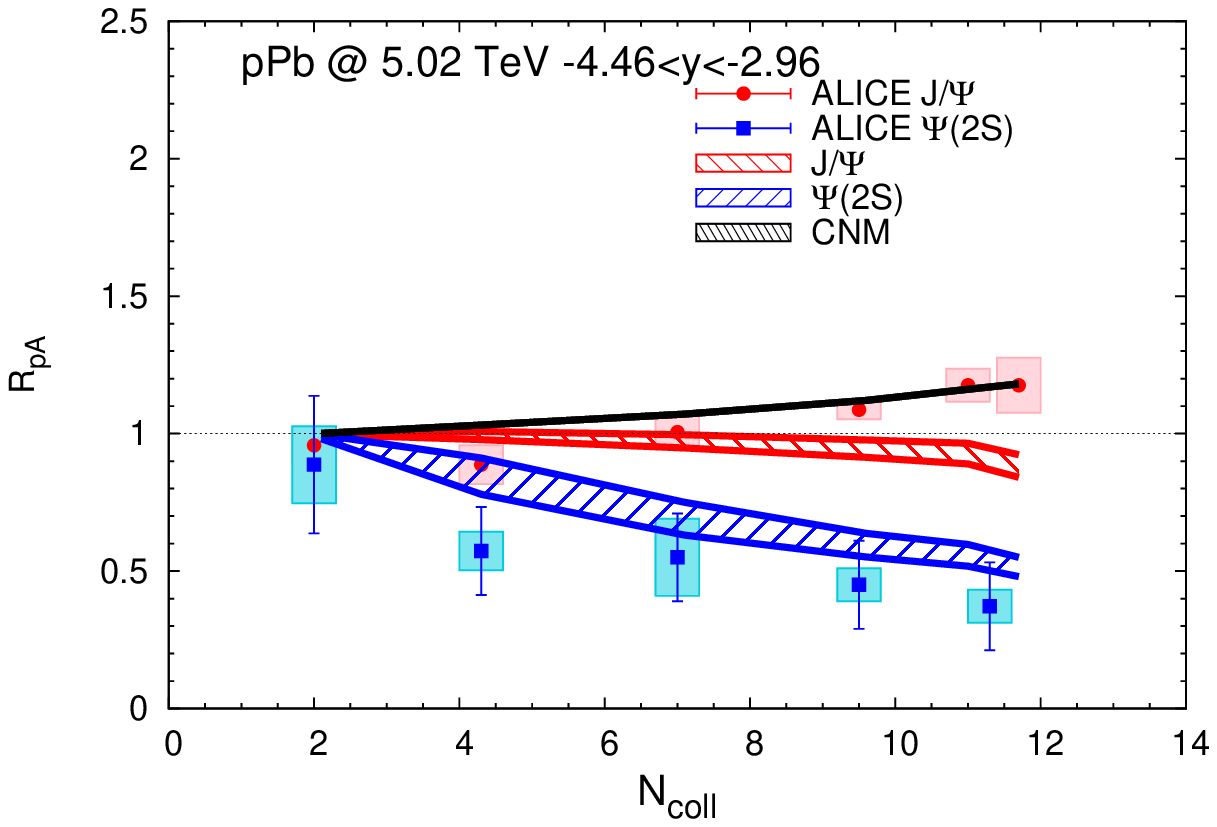}
\includegraphics[width=0.45\textwidth]{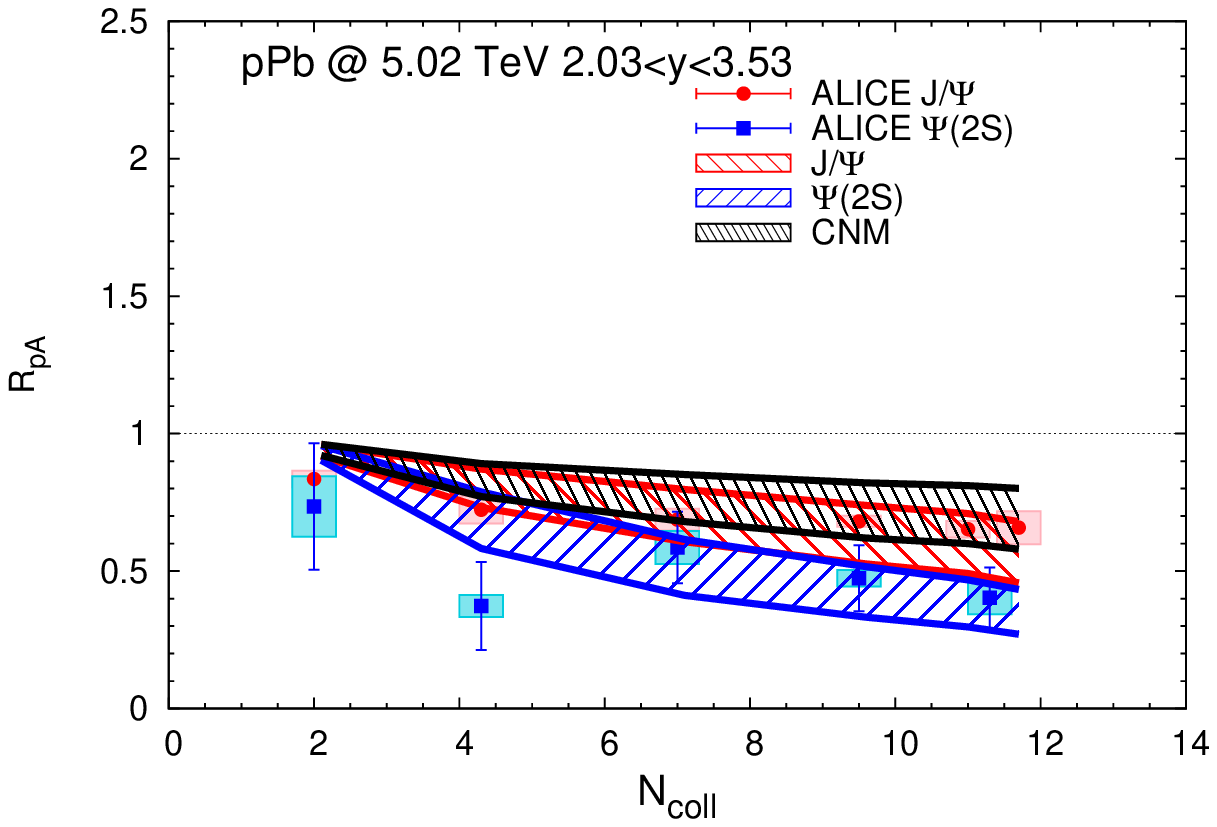}
\caption{Centrality-dependent $R_{p{\rm A}}$ for $J/\psi$ (red bands) and $\psi(2S)$
(blue bands) in 5.02\,TeV $p$-Pb collisions, compared with experimental
data~\cite{Adam:2015jsa,Leoncino:2016xwa,Adam:2016ohd}. The left (right) panel is for
backward (forward) rapidity. The bands are due to (anti-) shadowing from EPS09 LO/NLO~\cite{Vogt:2010aa,Vogt:2015uba}
at forward (backward) rapidity, as illustrated by the black bands which do not include
final-state effects.}
\label{raa_5020}
\end{figure}
\begin{figure}[!t]
\includegraphics[width=0.45\textwidth]{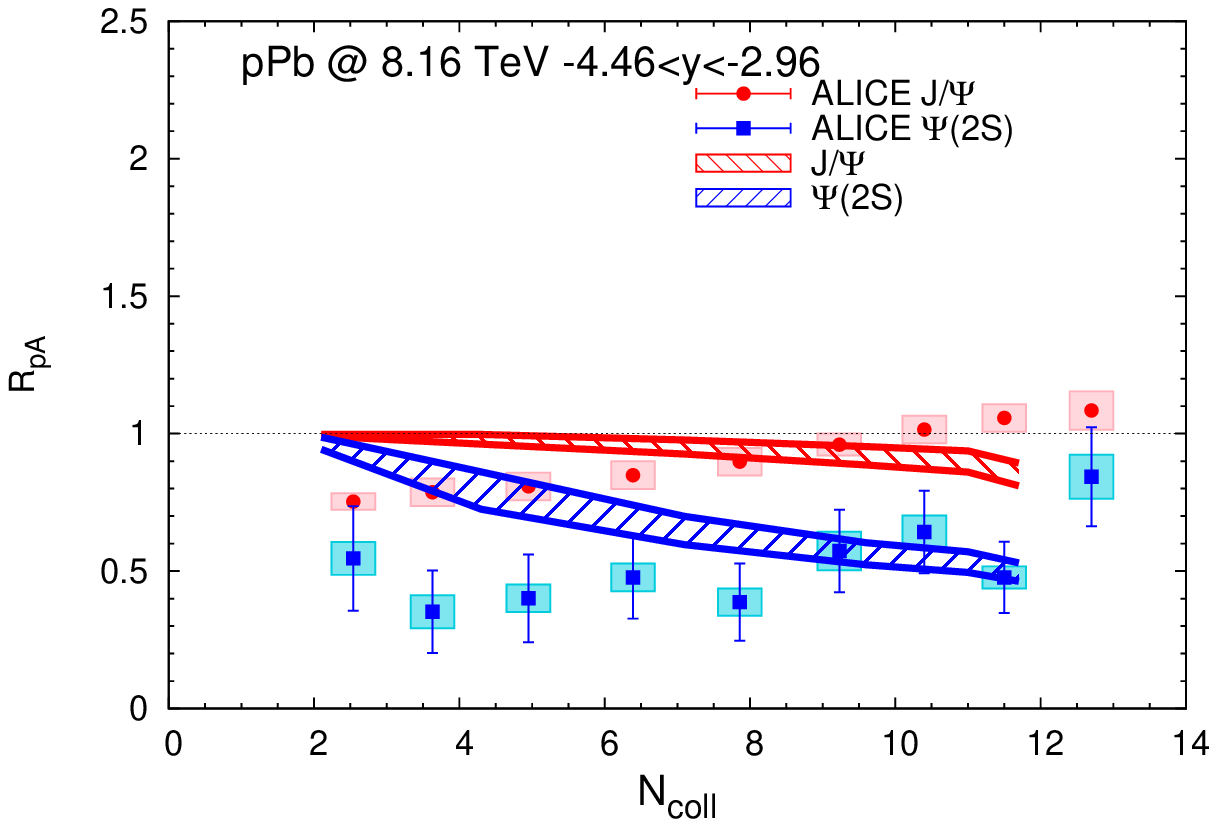}
\includegraphics[width=0.45\textwidth]{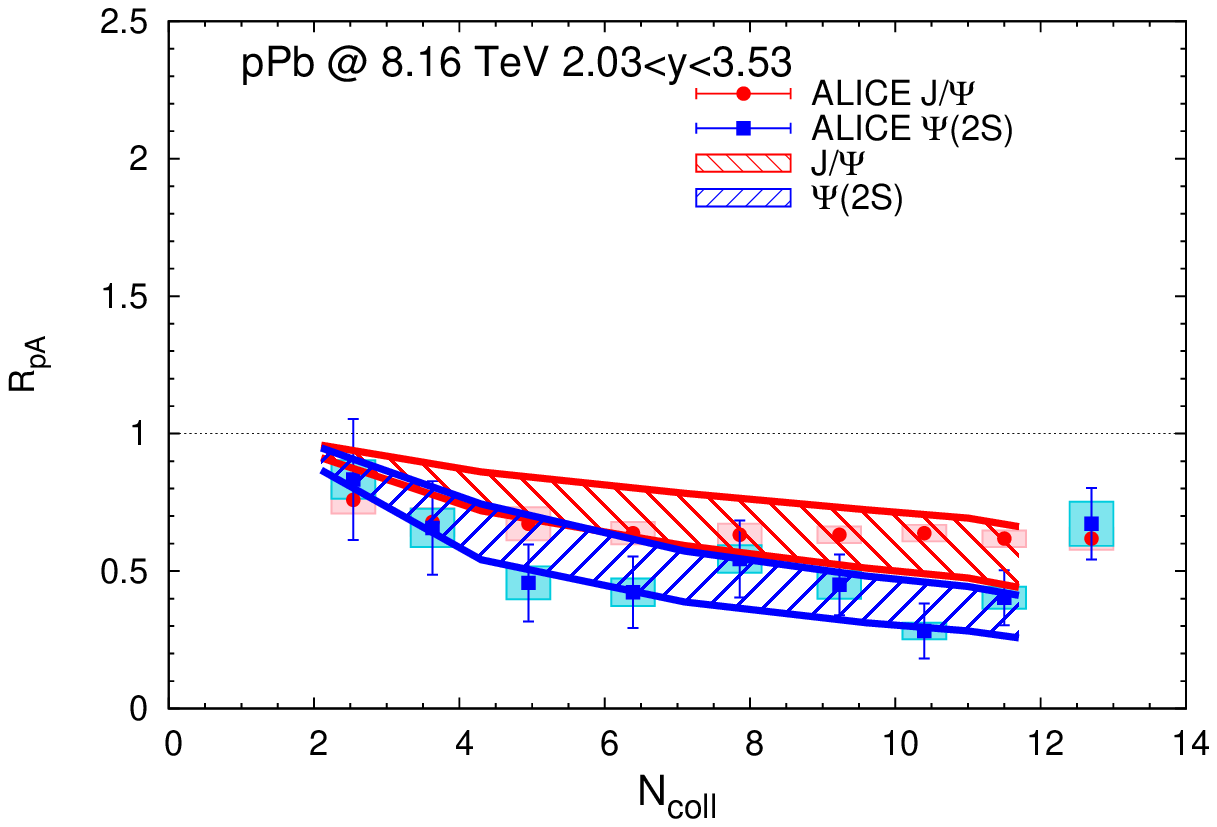}
\caption{Centrality-dependent $R_{p{\rm A}}$ for $J/\psi$ (red bands) and $\psi(2S)$
(blue bands) in 8.16\,TeV $p$-Pb collisions, compared with experimental
data~\cite{alice:ALICE-PUBLIC-2017-007}. The left (right) panel is for backward
(forward) rapidity. The bands are due to anti/-shadowing from an EPS09 LO/NLO~\cite{Vogt:2010aa,Vogt:2015uba} at forward (backward) rapidity.}
\label{raa_8160}
\end{figure}

In determining the centrality of a $p$-Pb collisions, we adopt the charged-particle
multiplicity and relate it to the average binary collision number following
Refs.~\cite{Adam:2015jsa,Adam:2014qja,Spousta:2014xja,Sirunyan:2017vpr}.
In Fig.~\ref{raa_5020}, our $J/\psi$ and $\psi(2S)$ calculations are compared to 5.02\,TeV
ALICE data. The black bands show only the CNM effects, bounded by the anti-/shadowing
obtained from EPS09-LO and EPS09-NLO calculations~\cite{Vogt:2010aa,Vogt:2015uba} for both
charmonia and open charm; as for the RHIC case, the centrality dependence of shadowing
is mimicked by a nuclear absorption-type behavior, while for anti-shadowing we
employ a parameterization of the pertinent lines shown in Fig.~3 of Ref.~\cite{Ferreiro:2014bia}.
The CNM effects dominate the uncertainty bands at forward rapidity (charmonium
formation time effects contribute $\sim$25\%); the uncertainty bands at backward
rapidity are entirely due to formation time effects (the same applies to Fig.~\ref{raa_8160}).
The shadowing-only bands already describe the $J/\psi$ data quite well. A moderate hot-matter
suppression of the $J/\psi$, together
with a small regeneration contribution of about 0.05 (in units of the $R_{p{\rm A}}$),
generate additional suppression which leads to a slight underestimation of the
backward-rapidity data but is compatible with the forward-rapidity data. For the $\psi(2S)$
the much larger suppression in the hot fireball is, however, essential to approximately
describe the suppression observed at both forward and backward rapidity.

In Fig.~\ref{raa_8160}, we compare our $J/\psi$ and $\psi(2S)$ calculations to 8.16\,TeV
ALICE data. There is a similar but slightly larger suppression compared to 5.02\,TeV for
both $J/\psi$ and $\psi(2S)$. We see quite some discrepancy with the data for peripheral
collisions at backward rapidity, but fair agreement with the data at forward rapidity.

\subsubsection{Transverse-Momentum Dependence}
\label{sssec_trans}
\begin{figure}[!t]
\includegraphics[width=0.45\textwidth]{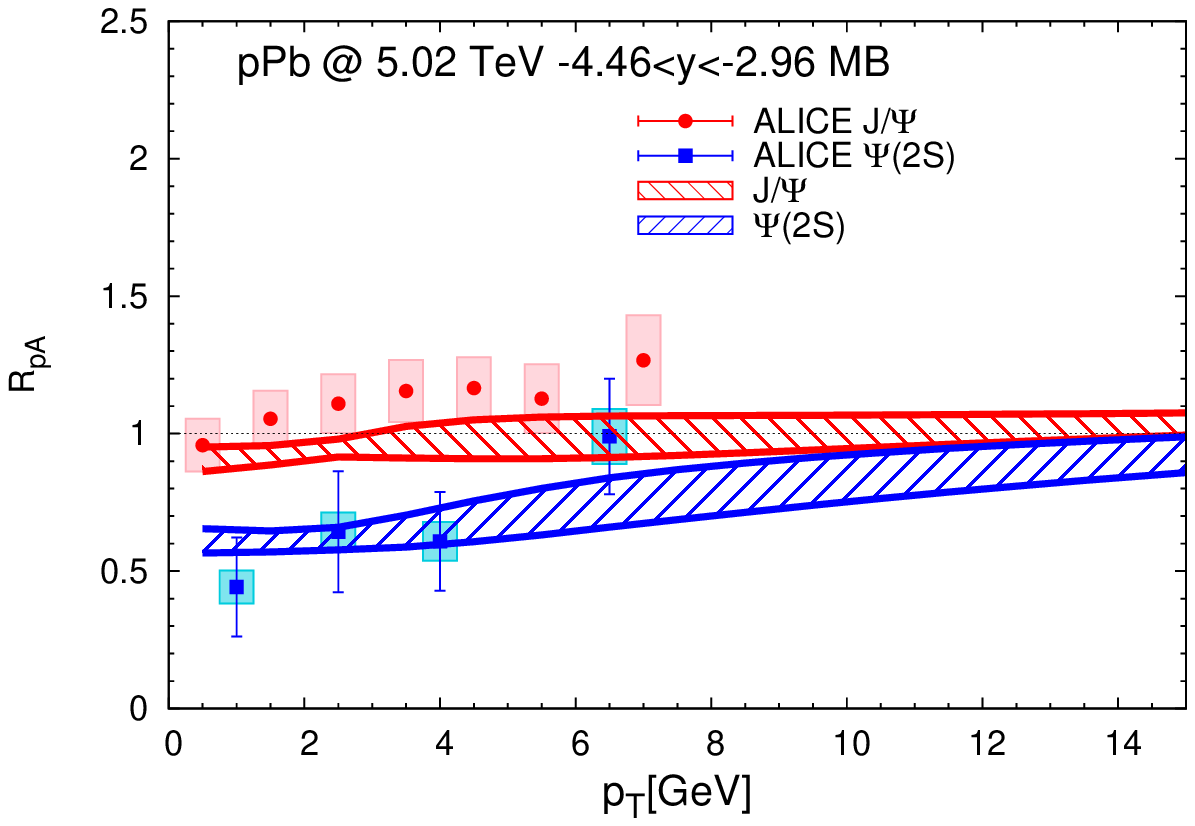}
\includegraphics[width=0.45\textwidth]{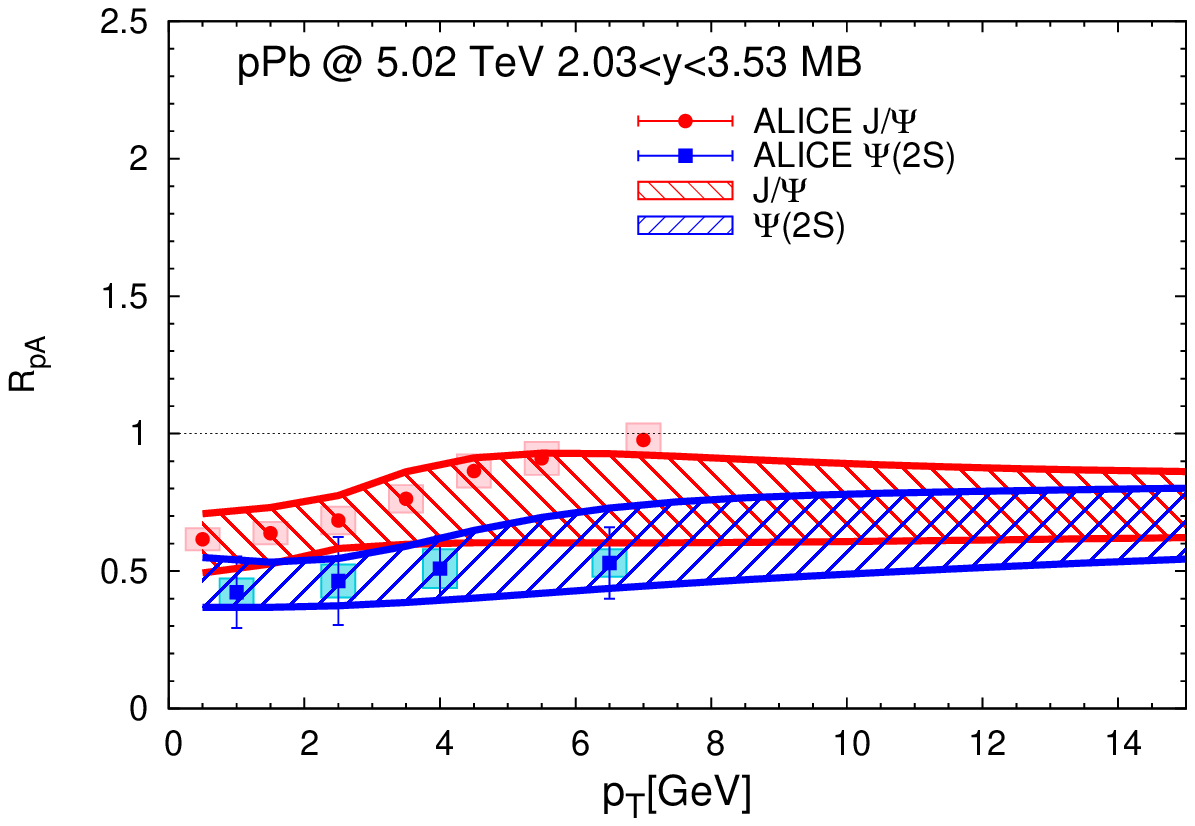}
\caption{Nuclear modification factor as a function of transverse momentum for $J/\psi$
(red bands) and $\psi(2S)$ (blue bands) in MB 5.02\,TeV $p$-Pb collisions, compared to
ALICE data~\cite{Adam:2015iga,Abelev:2014zpa}. The left (right) panel is for backward
(forward) rapidity. The uncertainty bands include variations in CNM and charmonium formation
time effects.}
\label{raa_502pT}
\end{figure}
\begin{figure}[!t]
\includegraphics[width=0.45\textwidth]{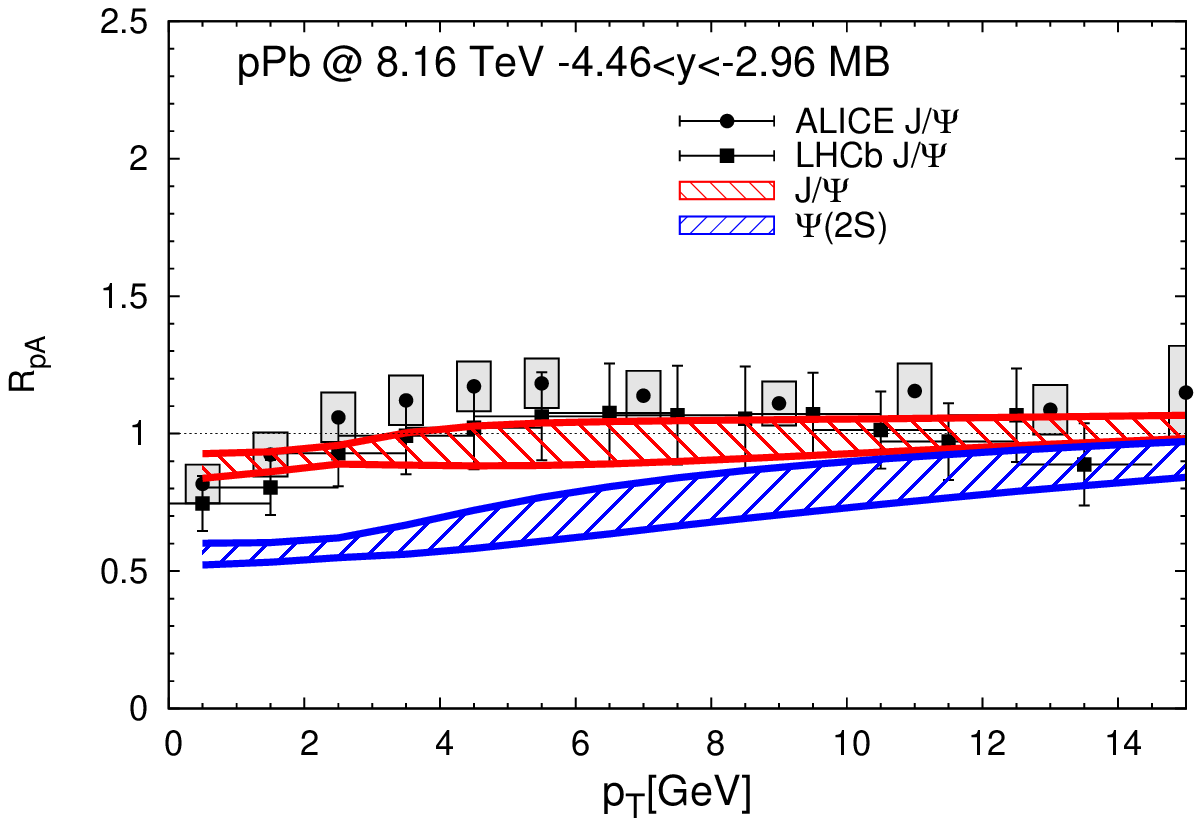}
\includegraphics[width=0.45\textwidth]{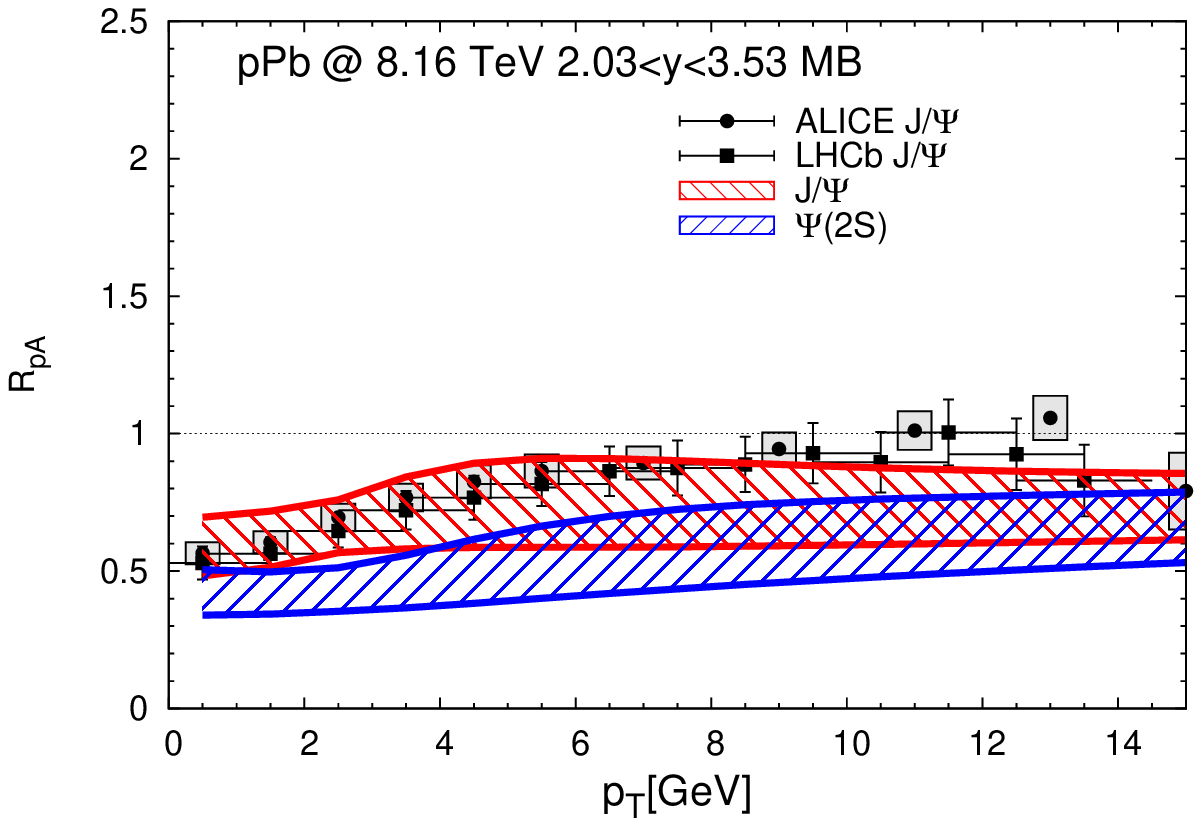}
\caption{Same as Fig.~\ref{raa_502pT} but for 8.16\,TeV $p$-Pb collisions with
ALICE and LHCb data~\cite{Acharya:2018kxc,Aaij:2017cqq} at 8.16\,TeV.}
\label{raa_8160pT}
\end{figure}
Our results for the $p_T$ dependence of charmonia, calculated as described in
Sec.~\ref{sec_rate}, are summarized in Figs.~\ref{raa_502pT} and \ref{raa_8160pT}
for minimum-bias (MB) $p$-Pb collisions at 5.02 and 8.16\,TeV, respectively.
We recall that an additional uncertainty arises through the $p_T$ dependence of shadowing,
which is incorporated into the theoretical bands conservatively as the maximum uncertainty
from all effects.

For the $J/\psi$, the calculated $R_{p{\rm A}}(p_T)$ at backward rapidities at both 5.02 and
8.16\,TeV exhibits a slight depletion at low $p_T$ followed by a mild maximum structure
around $p_T$$\simeq$5-6\,GeV, largely caused by the nuclear $p_T$ broadening. These trends become more pronounced at forward rapidity due to the generally increased
strength of the CNM effects.
Overall, the calculations are in agreement with ALICE and LHCb data within the theoretical
and experimental uncertainties at both energies.
The predictions for the $\psi(2S)$ reflect the stronger suppression already observed in the
centrality dependence. Relative to the $J/\psi$, most of the extra suppression is in
the low-$p_T$ region where the hot-matter effects are most pronounced while at higher
$p_T$, formation time effects mitigate the suppression.

\section{$J/\psi$ and $\psi(2S)$ Elliptic Flow}
\label{sec_v2}
\begin{figure}[!t]
\centering
\includegraphics[width=0.45\textwidth]{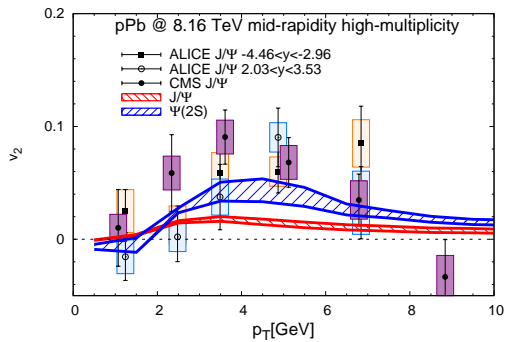}
\caption{Transverse-momentum dependent $v_2$ for $J/\psi$ (red band) and $\psi(2S)$
(blue band) at mid-rapidity in high-multiplicity $p$-Pb(8.16\,TeV) collisions within
the elliptic fireball model, compared to ALICE and CMS data~\cite{Acharya:2017tfn,CMS:2018xac}.}
\label{fig_v2}
\end{figure}

We finally turn to our calculation of the charmonium elliptic flow at 8.16\,TeV, where
data have recently become available~\cite{Acharya:2017tfn,CMS:2018xac}.
The primordial component acquires a positive $v_2$ from the path length differences
of the charmonium traversing the elliptic fireball, while the regeneration
component acquires its $v_2$ from the anisotropic flow in a blastwave description.
The primordial $v_2$ typically acquires values of 1-2\%, while the $v_2$ of
the regeneration component is much larger. However, since the latter, as mentioned
above, is limited to $R_{p{\rm A}}$ contributions of around 0.05-0.10, its weight in
the total $v_2$ is small. Our results shown in Fig.~\ref{fig_v2} predict a small
$v_2$ of up to $\sim$2\% for the $J/\psi$, and a larger value of up to $\sim$5\%
for the $\psi(2S)$, in high-multiplicity (most central) $p$-A collisions. We have
checked tested that the maximal $J/\psi$ $v_2$ generated from different versions of
the fireball parametrization does not exceed 2\%, essentially limited  by the constraints
from the initial eccentricity and the light-hadron $v_2$.
The near-zero result for the predominantly
primordial component of the $J/\psi$ is a direct consequence of its small hot-matter
suppression (and regeneration):
if it does not interact significantly, it cannot sense the spatial (or momentum)
anisotropies in the fireball. This is also the reason why the $v_2$ of the
$\psi(2S)$ is much larger, since the hot medium effects on it are much larger.
Since our $J/\psi$ results clearly underestimate the experimental data, we must conclude
that the the observed $v_2$ cannot originate from final-state interactions alone. The similar $v_2$ at
backward and forward rapidities (which have rather different multiplicities) is also in
line with this conclusion. One last caveat we can think of are {\em elastic} interactions
of the $J/\psi$ (and $\psi(2S)$) in the expanding medium, which we have not accounted for.
Very little is known about such interactions, and, in principle, one does not expect
them to be large due to the parametrically smaller size of the $J/\psi$ compared to light
hadrons, while for the $\psi(2S)$, due to its small binding, almost any interaction
can lead to break-up.

\section{Conclusion}
\label{sec_conclusion}
In the present work, we have extended our transport approach for in-medium quarkonia in
heavy-ion collisions to calculate $J/\psi$ and $\psi(2S)$ production in small collision
systems at RHIC ($d$-Au) and the LHC ($p$-Pb). Cold-nuclear-matter effects estimated from
nuclear parton distribution functions are combined with final-state effects treated
within a rate-equation framework for an expanding fireball including dissociation and
regeneration reactions in the QGP and hadronic phase. Our calculations provide a generally
fair description of the measured centrality and transverse-momentum dependent nuclear
modification factors measured in different rapidity regions, which differ in their CNM
and hot-nuclear matter effects (some tension with data was found in the 8.16\,TeV
backward-rapidity $R_{p{\rm A}}(N_{\rm coll})$). This supports an interpretation where
the $J/\psi$ observables are mostly dominated by CNM effects while the loosely bound
$\psi(2S)$ is subject to substantial suppression in the hot fireballs with initial
temperatures of about 200-300\,MeV and lifetimes of up to 4\,fm.
We also investigated the elliptic flow of $J/\psi$ and $\psi(2S)$. In our setup, a nonzero
$v_2$ results entirely from final-state interactions in the elliptic fireball.
Since the final-state suppression (and regeneration) especially for the $J/\psi$ is small,
which is compatible with the small hot-matter effects on the $R_{p{\rm A}}$, the resulting
$v_2$ is also small, not more than 2\% (and larger, up to 5\%, for the $\psi(2S)$); this
disagrees with the large signal observed in the LHC data.
We are therefore forced to conclude that this signal must be in large part due to initial-state
(or pre-equilibrium) effects not included in our approach. This situation appears to be
part of a bigger picture where the nuclear modification factor of hadrons, \eg,
$D$-mesons, shows little deviation from one while the $v_2$ is appreciable.

\vspace{0.3cm}

\acknowledgments
We thank Shuai~Y.~F.~Liu and Qipeng Hu for useful discussions. This work is supported by the
U.~S.~National Science Foundation (NSF) through grant no.~PHY-1614484.


\begin{thebibliography}{99}

\bibitem{Eichten:1974af}
  E.~Eichten, K.~Gottfried, T.~Kinoshita, J.~B.~Kogut, K.~D.~Lane and T.~M.~Yan,
  Phys.\ Rev.\ Lett.\  {\bf 34} (1975) 369.
   Erratum: [Phys.\ Rev.\ Lett.\  {\bf 36} (1976) 1276].

\bibitem{Rapp:2008tf}
  R.~Rapp, D.~Blaschke and P.~Crochet,
  Prog.\ Part.\ Nucl.\ Phys.\  {\bf 65} (2010) 209.

\bibitem{BraunMunzinger:2009ih}
  P.~Braun-Munzinger and J.~Stachel,
  Landolt-B\"ornstein {\bf 23} (2010) 424.

\bibitem{Kluberg:2009wc}
  L.~Kluberg and H.~Satz,
  Landolt-B\"ornstein {\bf 23} (2010) 372.

\bibitem{Mocsy:2013syh}
  A.~Mocsy, P.~Petreczky and M.~Strickland,
  Int.\ J.\ Mod.\ Phys.\ A {\bf 28} (2013) 1340012.

\bibitem{Rapp:2017chc}
  R.~Rapp and X.~Du,
  Nucl.\ Phys.\ A {\bf 967} (2017) 216.


\bibitem{Arnaldi:2010ky}
  R.~Arnaldi {\it et al.} [NA60 Collaboration],
  Phys.\ Lett.\ B {\bf 706} (2012) 263.

\bibitem{Adare:2013ezl}
  A.~Adare {\it et al.} [PHENIX Collaboration],
  Phys.\ Rev.\ Lett.\  {\bf 111} (2013) no. 20, 202301.

\bibitem{Adare:2016psx}
  A.~Adare {\it et al.} [PHENIX Collaboration],
  Phys.\ Rev.\ C {\bf 95} (2017) no. 3, 034904.

\bibitem{Sirunyan:2017mzd}
  A.~M.~Sirunyan {\it et al.} [CMS Collaboration],
  Eur.\ Phys.\ J.\ C {\bf 77} (2017) no. 4, 269.

\bibitem{Aaboud:2017cif}
  M.~Aaboud {\it et al.} [ATLAS Collaboration],
  Eur.\ Phys.\ J.\ C {\bf 78} (2018) no. 3, 171.

\bibitem{Sirunyan:2018pse}
  A.~M.~Sirunyan {\it et al.} [CMS Collaboration],
  arXiv:1805.02248 [hep-ex].

\bibitem{Aaboud:2018quy}
  M.~Aaboud {\it et al.} [ATLAS Collaboration],
  arXiv:1805.04077 [nucl-ex].


\bibitem{Adam:2015jsa}
  J.~Adam {\it et al.} [ALICE Collaboration],
  JHEP {\bf 1511} (2015) 127.

\bibitem{Leoncino:2016xwa}
  M.~Leoncino [ALICE Collaboration],
  Nucl.\ Phys.\ A {\bf 956} (2016) 689.

\bibitem{Adam:2016ohd}
  J.~Adam {\it et al.} [ALICE Collaboration],
  JHEP {\bf 1606} (2016) 050.

\bibitem{alice:ALICE-PUBLIC-2017-007}
  ALICE Collaboration [ALICE Collaboration],
  ALICE-PUBLIC-2017-007.

\bibitem{Adam:2015iga}
  J.~Adam {\it et al.} [ALICE Collaboration],
  JHEP {\bf 1506} (2015) 055.

\bibitem{Abelev:2014zpa}
  B.~B.~Abelev {\it et al.} [ALICE Collaboration],
  JHEP {\bf 1412} (2014) 073.

\bibitem{Acharya:2018kxc}
  S.~Acharya {\it et al.} [ALICE Collaboration],
  arXiv:1805.04381 [nucl-ex].

\bibitem{Aaij:2017cqq}
  R.~Aaij {\it et al.} [LHCb Collaboration],
  Phys.\ Lett.\ B {\bf 774} (2017) 159.

\bibitem{Acharya:2017tfn}
  S.~Acharya {\it et al.} [ALICE Collaboration],
  Phys.\ Lett.\ B {\bf 780} (2018) 7.

\bibitem{CMS:2018xac}
  CMS Collaboration [CMS Collaboration],
  CMS-PAS-HIN-18-010.

\bibitem{Albacete:2017qng}
  J.~L.~Albacete {\it et al.},
  Nucl.\ Phys.\ A {\bf 972} (2018) 18.

\bibitem{Ducloue:2016pqr}
  B.~Ducloué, T.~Lappi and H.~Mäntysaari,
  Phys.\ Rev.\ D {\bf 94} (2016) no. 7, 074031.

\bibitem{Arleo:2014oha}
  F.~Arleo and S.~Peigné,
  JHEP {\bf 1410} (2014) 073.
\bibitem{Ferreiro:2014bia}
  E.~G.~Ferreiro,
  Phys.\ Lett.\ B {\bf 749} (2015) 98.

\bibitem{Liu:2013via}
  Y.~Liu, C.~M.~Ko and T.~Song,
  Phys.\ Lett.\ B {\bf 728} (2014) 437.

\bibitem{Chen:2016dke}
  B.~Chen, T.~Guo, Y.~Liu and P.~Zhuang,
  Phys.\ Lett.\ B {\bf 765} (2017) 323.

\bibitem{Du:2015wha}
  X.~Du and R.~Rapp,
  Nucl.\ Phys.\ A {\bf 943} (2015) 147.

\bibitem{Nagle:2018}
  J.~Nagle, priv. comm. (2018).

\bibitem{Grandchamp:2003uw}
  L.~Grandchamp, R.~Rapp and G.~E.~Brown,
  Phys.\ Rev.\ Lett.\  {\bf 92} (2004) 212301.

\bibitem{Zhao:2010nk}
  X.~Zhao and R.~Rapp,
  Phys.\ Rev.\ C {\bf 82} (2010) 064905.

\bibitem{Grandchamp:2002wp}
  L.~Grandchamp and R.~Rapp,
  Nucl.\ Phys.\ A {\bf 709} (2002) 415.

\bibitem{Riek:2010fk}
  F.~Riek and R.~Rapp,
  Phys.\ Rev.\ C {\bf 82} (2010) 035201

\bibitem{He:2014cla}
  M.~He, R.~J.~Fries and R.~Rapp,
  Phys.\ Lett.\ B {\bf 735} (2014) 445.

\bibitem{Lin:1999ad}
  Z.~w.~Lin and C.~M.~Ko,
  Phys.\ Rev.\ C {\bf 62} (2000) 034903.

\bibitem{Liu:2017qah}
  S.~Y.~F.~Liu and R.~Rapp,
  Phys.\ Rev.\ C {\bf 97} (2018) 034918

\bibitem{Du:2017qkv}
  X.~Du, R.~Rapp and M.~He,
  Phys.\ Rev.\ C {\bf 96} (2017) 054901.

\bibitem{Kalaydzhyan:2014lja}
  T.~Kalaydzhyan and E.~Shuryak,
  Nucl.\ Phys.\ A {\bf 931} (2014) 899.

\bibitem{Vazquez:2017jce}
  O.~Vazquez [ALICE Collaboration],
  arXiv:1710.04715 [hep-ex].

\bibitem{Huovinen:2001cy}
  P.~Huovinen, P.~F.~Kolb, U.~W.~Heinz, P.~V.~Ruuskanen and S.~A.~Voloshin,
  Phys.\ Lett.\ B {\bf 503} (2001) 58.

\bibitem{Retiere:2003kf}
  F.~Retiere and M.~A.~Lisa,
  Phys.\ Rev.\ C {\bf 70} (2004) 044907.

\bibitem{Bossu:2011qe}
  F.~Bossu, Z.~C.~del Valle, A.~de Falco, M.~Gagliardi, S.~Grigoryan and G.~Martinez Garcia,
  arXiv:1103.2394 [nucl-ex].

\bibitem{Book:2015}
  J.~Book, PhD Thesis,
  University of Frankfurt (2015).

\bibitem{Eskola:2009uj}
  K.~J.~Eskola, H.~Paukkunen and C.~A.~Salgado,
  JHEP {\bf 0904} (2009) 065.

\bibitem{Vogt:2010aa}
  R.~Vogt,
  Phys.\ Rev.\ C {\bf 81} (2010) 044903.

\bibitem{Vogt:2015uba}
  R.~Vogt,
  Phys.\ Rev.\ C {\bf 92} (2015) no.3, 034909.

\bibitem{Hufner:2001tg}
  J.~H\"ufner and P.~Zhuang,
  Phys.\ Lett.\ B {\bf 515} (2001) 115.

\bibitem{Ducloue:2015gfa}
  B.~Ducloué, T.~Lappi and H.~Mäntysaari,
  Phys.\ Rev.\ D {\bf 91} (2015) no.11, 114005.

\bibitem{Adare:2011vq}
  A.~Adare {\it et al.} [PHENIX Collaboration],
  Phys.\ Rev.\ D {\bf 85} (2012) 092004

\bibitem{Adare:2010de}
  A.~Adare {\it et al.} [PHENIX Collaboration],
  Phys.\ Rev.\ C {\bf 84} (2011) 044905

\bibitem{Adamczyk:2012af}
  L.~Adamczyk {\it et al.} [STAR Collaboration],
  Phys.\ Rev.\ D {\bf 86} (2012) 072013

\bibitem{Adam:2016rdg}
  J.~Adam {\it et al.} [ALICE Collaboration],
  Phys.\ Lett.\ B {\bf 766} (2017) 212.

\bibitem{Acharya:2017hjh}
  S.~Acharya {\it et al.} [ALICE Collaboration],
  Eur.\ Phys.\ J.\ C {\bf 77} (2017) no.6, 392.

\bibitem{Zhao:2011cv}
  X.~Zhao and R.~Rapp,
  Nucl.\ Phys.\ A {\bf 859} (2011) 114.

\bibitem{Adam:2014qja}
  J.~Adam {\it et al.} [ALICE Collaboration],
  Phys.\ Rev.\ C {\bf 91} (2015) no.6, 064905.

\bibitem{Spousta:2014xja}
  M.~Spousta [ATLAS Collaboration],
  Nucl.\ Phys.\ A {\bf 932} (2014) 404.

\bibitem{Sirunyan:2017vpr}
  A.~M.~Sirunyan {\it et al.} [CMS Collaboration],
  JHEP {\bf 1801} (2018) 045.


\end{thebibliography}
\end{document}